# Nanosensitive optical coherence tomography to assess wound healing within the cornea

**CERINE LAL[1], SERGEY ALEXANDROV[1], SWETA RANI[2], Yi ZHOU[1], THOMAS RITTER[2], MARTIN LEAHY[1*]**

[1] *Tissue Optics and Microcirculation Imaging Facility, National Biophotonics and Imaging Platform, School of Physics, National University of Ireland, Galway*
[2] *Regenerative Medicine Institute, National University of Ireland, Galway, Ireland*
[*]*martin.leahy@nuigalway.ie*

**Abstract:** Optical Coherence Tomography (OCT) is a non-invasive depth resolved optical imaging modality, that enables high resolution, cross-sectional imaging in biological tissues and materials at clinically relevant depths. Though OCT offers high resolution imaging, the best ultra-high-resolution OCT systems are limited to imaging structural changes with a resolution of one micron *on a single B-scan* within very limited depth. Nanosensitive OCT (nsOCT) is a recently developed technique that is capable of providing enhanced sensitivity of OCT to structural changes. Improving the sensitivity of OCT to detect structural changes at the nanoscale level, to a depth typical for conventional OCT, could potentially improve the diagnostic capability of OCT in medical applications. In this paper, we demonstrate the capability of nsOCT to detect structural changes deep in the rat cornea following superficial corneal injury.

## 1. Introduction

OCT is a low coherence interferometric imaging technique that maps axial reflections of near-infrared (NIR) light from tissue to form cross sectional images of morphological features at the micrometer scale [1,2]. Since its introduction early 1990's, OCT has been clinically demonstrated in a diverse set of medical and surgical applications, including ophthalmology, gastroenterology, dermatology, cardiology, and oncology, among others [1]. Being a non-invasive imaging modality, OCT can be used to produce cross sectional images of tissues in situ and in real time, without the need to excise and process the specimens, as in conventional biopsy and histopathology procedures. The penetration depth of OCT is limited by the optical scattering and is up to 2–3 mm in biological tissues. Although this depth is shallow compared with other clinical imaging techniques, the image resolution of OCT both lateral and axial, is 10–100 times finer than conventional 3D imaging techniques such as the ultrasound, magnetic resonance imaging and computed tomography [2]. In a conventional OCT system, the lateral resolution is inversely proportional to the numerical aperture of the objective lens and the axial resolution is limited by the bandwidth of light sources used for imaging [1]. Typical values for the axial resolution are 6–15 μm in air. Over the years, numerous techniques have been proposed and implemented to improve the axial resolution of conventional OCT systems. Most of these techniques used Ti:sapphire lasers or light sources based on supercontinuum generation from photonic crystal fibres for imaging [3–5]. However, the best ultra-high resolution OCT systems are limited to detecting structural changes with a resolution of one micron *on a single B-scan* within very limited depth [3]. Currently, the axial resolution for imaging in scattering tissues is generally limited to a maximum of (imaging depth)/200 [6]. Though several super-resolution and nanoscale detection techniques were proposed recently, imaging structures over four orders of magnitude in size, using the same imaging technique, remains a challenge [7].

OCT is used for both structural and functional imaging of *in vivo* biological tissues. The functional imaging applications include angiographic OCT, photothermal OCT, spectroscopic OCT and elastography. However, most of these functional imaging techniques are unable to determine submicron changes within the tissue and are capable of only determining the structural/functional aspects within the optical resolution of the OCT system, excluding spectroscopic OCT and phase sensitive methods. Recently, a few methods have been proposed in combination with OCT to detect nanoscale structural changes. One of them is based on phase sensitive OCT (psOCT) which uses the Fourier analysis of phase differences of acquired interference spectra ( A- scans) at particular depth positions [8]. Nanoscale detection with phase sensitive techniques has been used for different applications, including optical coherence elastography (OCE) to detect the biomechanical properties [9] , to determine the submicron movement of the basilar membrane within the organ of Corti and neural action potential in a squid and others [10,11]. However, phase sensitive OCT requires two or more frames (M-scan) for realization and is inherently prone to noises (bulk motion, vibrations, etc.). Another technique called inverse spectroscopic OCT (ISOCT) has been developed wherein wavelength dependent backscattering coefficient and scattering coefficient is determined by fitting an autocorrelation function to the detected A-scan OCT signal and from the fit model based on Mie scattering theory, the mass density distribution in a biological sample is quantified [12]. The mass density distribution obtained using ISOCT was used to study extra cellular matrix remodelling in *in vitro* cancer models [13] and to study field carcinogenesis [14].

Recently, nano-sensitive OCT has been developed by Alexandrov *et al* [15–18]. It is based on the spectral encoding of spatial frequency approach [18–21] which demonstrated nanoscale sensitivity to structural changes and super-resolution imaging. The nsOCT has been used to detect both structural and dynamic changes in *ex vivo* and *in vivo* biological tissues [15,16,23]. The nsOCT permits access to the local spatial frequency content of the object directly based on the general scattering theory [24]. Hence, nsOCT provides quantitative information about structural sizes within the accessible range of spatial frequencies. The length scale in nsOCT depends on the spectral bandwidth of the light source and is in the sub-micron scale even for 1300 nm central wavelength source (from ~620 nm to ~680 nm optical length scale in the present paper). However, sensitivity to structural alterations in time and in space is at nanoscale, as it was shown in references [15,16,23]. Also, spatial frequency domain correlation mapping optical coherence tomography has been described recently [25] , for detection of depth resolved nanoscale structural changes non-invasively based on the principles of nano-sensitive OCT.

Studying nanoscale structural and dynamic changes *in vivo* is fundamental to understanding changes occurring at cellular level before the changes manifest at the tissue level. Detecting these submicron structural changes can help scientists and clinicians to diagnose the onset of a disease, its progression and in determining treatment effectiveness of drugs. Herein, OCT offers great potential whereby combining nano-detection techniques together with its real time, 3D structural imaging capability, can provide sub voxel structural data by mapping nanoscale structural changes without improving the actual optical resolution. In contrast to phase sensitive OCT, nsOCT images are less sensitive to noise and can be formed using just one frame. Both techniques detect different information: nsOCT provides information about structural changes whereas psOCT detects the displacements in time, and so can be complementary to each other. nsOCT can partially overcome the scale range issue in optical imaging modalities, and also is cost effective without the need to use expensive high resolution imaging optics.

The cornea is the transparent, avascular layer of the eye that controls the entry of light into the eye and helps to refract the light onto the retina. Corneal transparency is vital to preserve its structure and function. Corneal injuries generally arise from thermal and chemical burns [26,27]. Of these, 11.5 – 22% of all ocular injuries occur from chemical burns, from both acids and alkali [28]. Among chemical induced corneal burns, alkali burn causes more damage to the corneal stroma and anterior chamber compared to acid injury. Alkali ions being lipophilic, penetrate into the corneal stroma disrupting the cells and denaturing the collagen matrix, which promotes further penetration into the anterior chamber [29,30]. The corneal stroma plays a vital role in maintaining corneal transparency and acts as a load - bearing agent by protecting the ocular tissues from changes in intra ocular pressure. Any change within the micro structures of the cornea results in loss of transparency and increases the light scattering [31]. The corneal stroma is made up of collagens, proteoglycans, glycoproteins and keratocytes and it has been shown that, it is the nanoscale arrangement of collagen fibrils that ensures corneal transparency. It has been reported that any change in the diameter of the collagen fibrils or creation of voids between the fibrils causes increased light scattering within the cornea and leads to corneal opacity [31]. Another factor that increases light scattering within the cornea is the activation of keratocytes within the stroma in response to the corneal wound healing process. Based on these studies, it is imperative to understand the nanoscale structural changes occurring during ocular injury and subsequent wound healing process *in vivo* for assessment of wound repair and monitoring treatment efficacy.

Over the years, OCT has been routinely used in ophthalmic applications and some recent studies have reported its use in evaluating the chemical ocular burn and its healing process [32–34]. All of these studies were based on the analysis of the structure of the anterior segment of the injured eye from the OCT B-scans. As these chemical agents alter the structural integrity of the cornea upon contact, nsOCT provides the possibility to detect and visualize the sub-micron structure from just one frame and the submicron changes using two frames which otherwise cannot be obtained from conventional OCT and other images.

In this paper, we investigate the applicability of nsOCT to detect these submicron structural changes within the cornea following superficial alkali injury in a pre-clinical rat model. The results obtained by nsOCT are validated with results from corneal histology sections.

## 2. Materials and methods

*2.1 Experimental set up:*

In the present study, for pre-clinical imaging, a commercial VCSEL based swept source OCT system operating at 200 kHz (OCS1310V2, Thorlabs) was used. The system was operating at a central wavelength of 1300 nm (source bandwidth of 117 nm) with longest and shortest wavelength of the source being 1358 nm and 1241 nm respectively. The system had an axial resolution of 16 µm in air specified by the manufacturer. For the study, we used 5X objective (LSM03, Thorlabs) that provided a spatial resolution of 25 µm. The average output power measured in the sample arm was 5 mW and had a signal to noise ratio (SNR) of 98 dB.

*2.2 Rat cornea alkali burn model:*

Application of alkali to one cornea of the rat was performed under anaesthesia with isoflurane followed by topical tetracaine. To induce alkaline injury, a piece of Whatman filter paper (3mm diameter) was soaked in NaOH (4 µl of a 1 M solution) and applied to the centre of the cornea of the right eye for 60 seconds followed by rinsing with 10 ml of saline for fifteen minutes. Male Lewis rats aged 8-14 weeks were obtained from Harlan Laboratories UK and were housed with food and water for the study in a fully accredited animal housing facility. This study was approved by Animals Care Research Ethics Committee of the National University of Ireland, Galway. All the experimental procedures were performed in accordance with and authorization from the Health Products Regulatory Authority of Ireland.

For histology analysis, animals were euthanized on the $7^{th}$ day and the intact enucleated eyes were fixed in 10% neutral buffered formalin and paraffin embedded using the Leica ASP300 tissue processor. Paraffin-embedded eyes were then sectioned (5 µm; Leica Microtome) and deparaffinised by sequential washing with xylene followed by washing in a descending series of ethanol and stained using haematoxylin and eosin (Sigma-Aldrich). The stained sections were examined by Olympus light microscopy (20X magnification).

*2.3 Experimental design*

For the OCT imaging, the rats were mounted on an in-house developed mounting system such that the eyes were stabilized and caused minimal movement artefact. OCT interferograms were acquired prior to alkali injury and after the injury (time, $t$ =0) of the study. To assess the wound healing process, subsequent imaging was performed on day 7 following the alkali injury. Twelve rats were included in this study, and raw OCT spectral interference signals were recorded before (termed as the healthy group) and after inducing the alkali injury (termed as the immediate phase group). Five rats from the immediate phase group were imaged on the 7th day following the injury (termed as the acute reparative phase).

The mean optical power at the output of the sample arm was 5mW, which is well below the American National Standards Institute limit for maximum permissible exposure of 15.4 mW at the wavelength of operation. For all the imaging, 3D volumetric data were acquired covering an area of 5×5 mm$^2$, with 200 B - scans covering the entire scan range.

*2.4 nsOCT image formation*

Nano-sensitive OCT accesses the three-dimensional spatial frequency components of a sample, as a two-dimensional spatial frequency distribution in the Fourier plane [15–18]. It accomplishes this by measuring the wavelength distribution for each voxel in a Fourier domain OCT system. In a typical Fourier domain OCT system, the measured Fourier components ($v_z$) of the backscattered signal on the spectrometer at reasonably small numerical aperture (*NA*) of the objective lens is given by Eq. (1), where $\lambda$ is the central wavelength of the source, $n$ is the refractive index and the corresponding spatial frequency period ($H_z$) is given by Eq. (2).

$$v_z = \frac{2n}{\lambda} \qquad (1)$$

$$H_z = \frac{1}{v_z} \qquad (2)$$

Depending on the source bandwidth, there exists a range of spatial frequencies that can be captured on the detector which is given by Eq. (3), where $\lambda_2$ and $\lambda_1$ are the longest and shortest wavelength of the source and $\Delta\lambda$ is the bandwidth of the source.

$$\Delta v = \frac{2n\Delta\lambda}{\lambda_1\lambda_2} \qquad (3)$$

The scattered waves for a given wavelength will be in phase only if the spacing between the reflected planes is equal to one half of the wavelength. This implies that, for a given spacing or a given spatial period, a strong signal is detected only at one wavelength. In OCT, since the directions of illumination and measurement of backscattering are the same, the spatial frequency/period can be obtained by Eq. (1) and Eq. (2) respectively. From Eq. (1) and (2), we can observe that a change in spatial periodicity of the structure by $\Delta H_z$, results in a wavelength shift $\Delta\lambda = 2n\Delta H_z$. Such a shift can be easily detected by an OCT system (spectral resolution of our OCT system is 0.093 nm). However, while taking the inverse Fourier transform of the interference signal to reconstruct the OCT structural image, the spatial frequency information which corresponds to small, submicron structure, is lost. This reduces the sensitivity of conventional OCT signal processing to detect submicron changes in the scattering structures. In nsOCT, by scaling each of the spatial frequencies, or spatial frequency periods, to the corresponding wavelength, the spatial frequency of the scattering structures is preserved when transforming from *k*-space to the image space, thereby enhancing the sensitivity of OCT imaging.

In order to realize nsOCT, first the *k*-space linearized spectral interferogram $I(\lambda)$ is converted to corresponding axial spatial frequency $I(v_z)$ using Eq. (1). In the present study, $\Delta\lambda$ is 117 nm corresponding to $\lambda_1$ and $\lambda_2$ of 1241 nm and 1358 nm respectively. The spatial frequencies varied from 2.025 MHz to 2.21 MHz according to Eq.1 ($n$ = 1.376 for cornea) and corresponding physical spatial periods varied from 451 nm to 493 nm according to Eq. 2. If we consider the refractive index, $n$ to be equal to 1, the optical spatial periods calculated according to Eq.1 will vary from 621 nm to 679 nm. In this paper, we have used physical spatial periods for our analysis.

The spectrum of axial spatial frequency is then decomposed into *N* zones using a Tukey window. For each of the *N* zones, the axial spatial frequency profile is inverse Fourier transformed to reconstruct the OCT image for each zone. From the reconstructed OCT images of *N* zones, the dominant spatial frequency/period at each point is determined by finding the maximum intensity values at each point across the *N* zones. Next, the dominant spatial frequency/ period value is mapped to form the nsOCT image. This process is repeated for every A-line in a B-scan and for every B-scan in a 3D volume. In the present study, the spectral interference signal was decomposed into 10 zones with each zone having spatial frequency bandwidth ($\delta v_N$) of $\frac{\Delta v}{10}$. There is a trade-off between spatial resolution and structural resolution depending on the width of the window used. We can apply different width of the window depending on the sample and purpose of imaging, and so improve structural or spatial resolution. To reconstruct nsOCT spatial period profiles, the windowed spectrum was inverse Fourier transformed using *p*-point IFFT where *p* is the length of the interference signal. Briefly, the technique is described in the flowchart shown in figure 1. Further, to supress the noise within the nsOCT images, a 4x4 spatial kernel and a threshold of mean + 0.7 *standard deviation was used. The spatial filtering kernel and the threshold can be optimized based on the application or requirement.

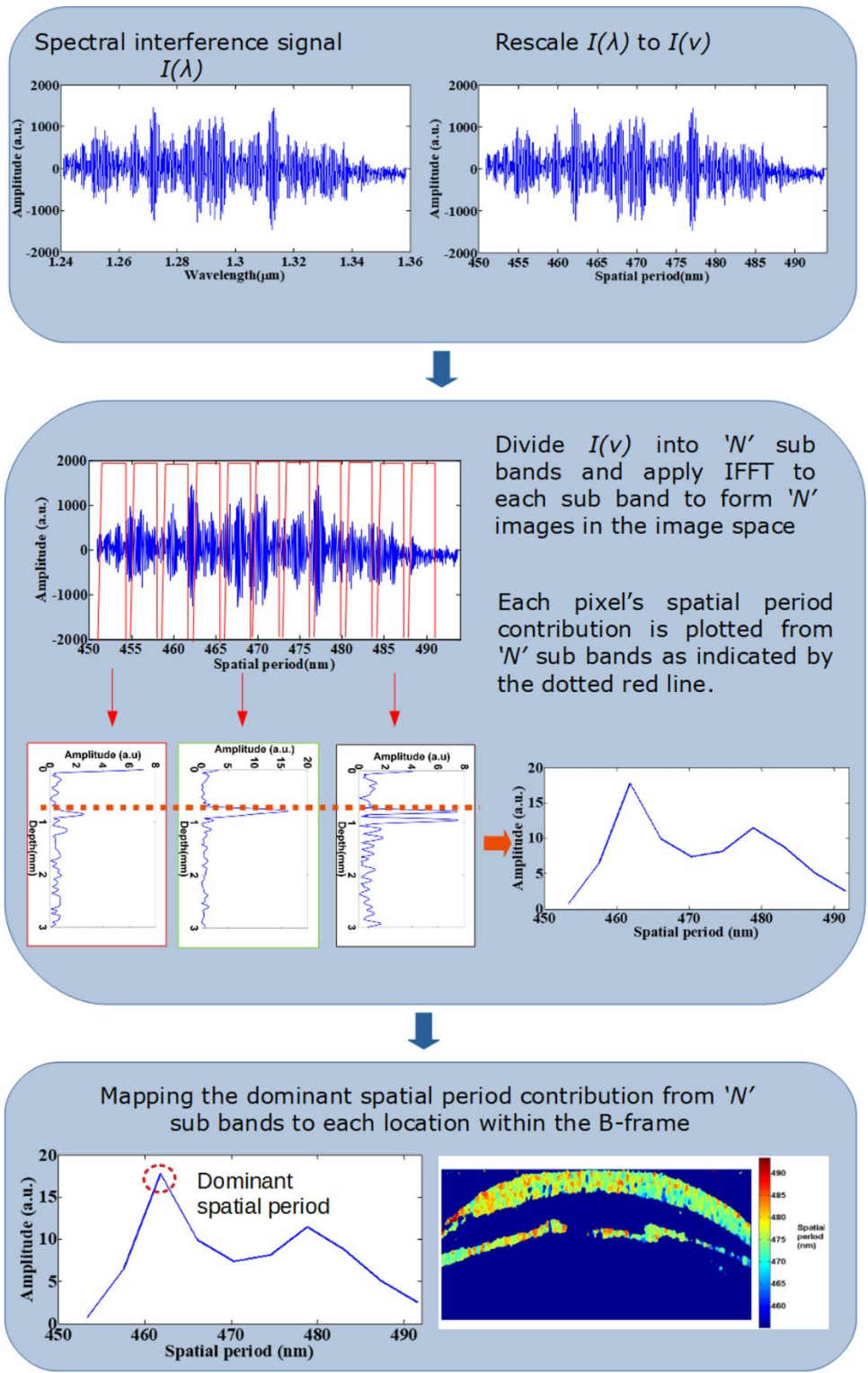

Fig. 1. Flow chart describing nsOCT image formation.

## 3. Results and discussion

To demonstrate the ability of nsOCT to detect depth resolved structural changes at nanoscale, we imaged two samples with periodic axial structure, Bragg gratings obtained from OptiGrate Corp. USA. Images of these samples with different and well known axial periodic structures (431.6 nm and 441.7 nm and a refractive index of 1.48 ± 0.001) are shown in figure 2. Figure 2(a) shows the conventional OCT B scan image obtained from the Bragg gratings and figure 2(b) shows the corresponding nsOCT spatial period maps.

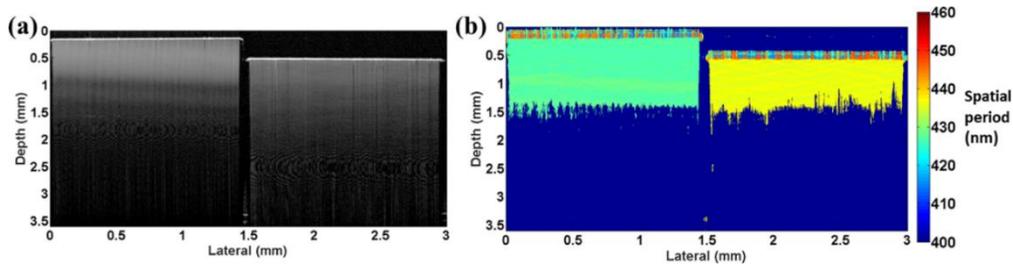

Fig. 2. Experimental demonstration of nsOCT technique using Bragg grating having axial spatial period of 431.6 nm and 441.7 nm (a) OCT B scan (b) nsOCT map

From the above nsOCT processed images of the Bragg gratings, it can be observed that our technique clearly detects the sub-micron axial spatial periods of the samples under investigation and is not detecting signals arising from any optical aberrations. The Bragg gratings has an antireflection coating on top and gives an appearance of noise signal as can be seen in figure 2(b). Also, figure 2(b) clearly demonstrates that chromatic aberrations of the imaging system versus depth are negligible and we can clearly visualise structures with different sizes at a depth of about 1 mm, and perhaps deeper. From figure 2(b), the depth resolved difference in structural size of 10 nm can be detected. Figure 2 confirms that using nsOCT technique, we can detect the sub-micron structure and nanoscale differences between such structures without resolving them spatially.

Next, for the assessment of corneal wound healing process in a pre-clinical rat burn model, nsOCT algorithm was implemented to determine the nanoscale structural changes occurring within the cornea over time. Figure 3 shows the conventional OCT B-scans and nsOCT B-scans of healthy rat cornea, cornea after alkali injury and the same cornea after 7 days.

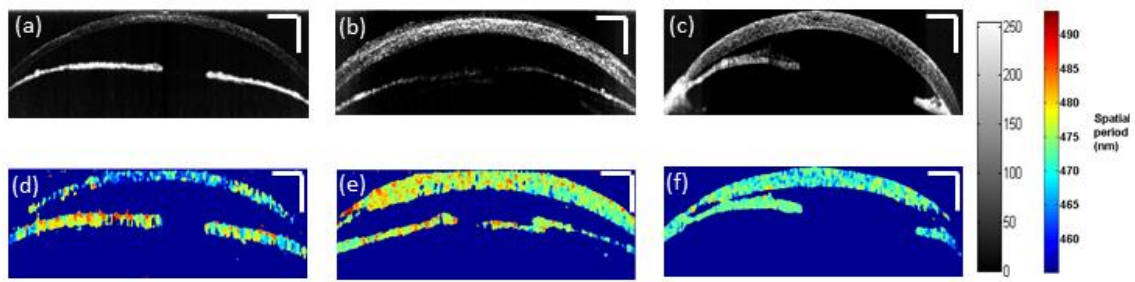

Fig. 3. Conventional OCT B-scans (a) uninjured cornea (b) after alkali induced corneal burn (c) injured cornea on 7[th] day (d) nsOCT B-scan of corresponding uninjured cornea (e) nsOCT B-scan following alkali induced corneal burn (f) nsOCT B-scan on 7th day following the injury. Scale bars – 500 µm.

Alkali burn results in corneal oedema and increased light scattering within the cornea as can be observed from OCT intensity B–scan images in figure 3(b) and 3(c), but there is no information about structural changes within the cornea. Also, from the conventional OCT intensity images, we can observe the thickening of the cornea following the alkali injury. Upon alkali burn, the average thickness of the central cornea increased from 150 ± 3.6 μm (mean ± standard error) for the healthy group to 270 ± 10 μm (mean ± standard error) for the immediate phase group and to 315 ± 39 μm (mean ± standard error) during the acute reparative phase. While reconstructing OCT intensity B-scans following the conventional approach, we lose the sensitivity of OCT to spatial period information. From figures 3(d) – (e) we can observe that nsOCT processed B-scans differ significantly in the spatial period of the structures within the cornea between a healthy cornea, at the onset of alkali injury and during the acute reparative phase of the injured cornea. The corneal inflammation and denaturization of the collagen matrix in response to alkali injury results in an increase in spatial period of the structures within the cornea as observed in nsOCT images in figure 3(e) and 3(f) compared to figure 3(d).

As stated before, alkali injury penetrates the corneal stroma and leads to the damage of anterior chamber. Hence, for better visualization of the nanoscale structural changes within the cornea at each depth, *enface* images were reconstructed from the processed nsOCT and conventional B-scans excluding the iris [35]. Figure 4 shows representative *enface* OCT images at a depth of 60 µm before and after the alkali injury. From the figure, it can be observed that conventional OCT *enface* intensity images fail to distinguish between healthy cornea and injured cornea. In conventional structural OCT images, including images in figures 4 a - c, the intensity value at each point provides information only about reflectivity at a given location and does not convey any information about the structure below the resolution limit at that location. To quantitatively assess if changes in OCT intensity alone could detect changes within the cornea, the intensity values across the corneal depth for all the three groups are plotted in figure 5. From figure 5, we can observe that intensity alone does not provide any information regarding the depth dependant structural changes occurring within the cornea following the alkali burn and subsequent healing. Hence, it is not possible to detect if the structure within one area is different from the structure within another area solely based on intensity/gray level values.

In contrast, nsOCT images are formed using a different contrast mechanism, i.e., they visualize the dominant size of the sub-micron structure at a given location. Thus, from nsOCT images, the structural changes can be detected as shown in figures 4 d – f. nsOCT processed *enface* images clearly indicate changes in spatial period within the cornea and helps to distinguish between healthy and injured cornea as shown in figure 4(d) and 4(e). Also, nsOCT processing is able to track structural changes happening within the cornea at the onset of an injury and also in assessing the healing process based on the changes in the spatial period as can be observed in figures 4(e) and 4(f).

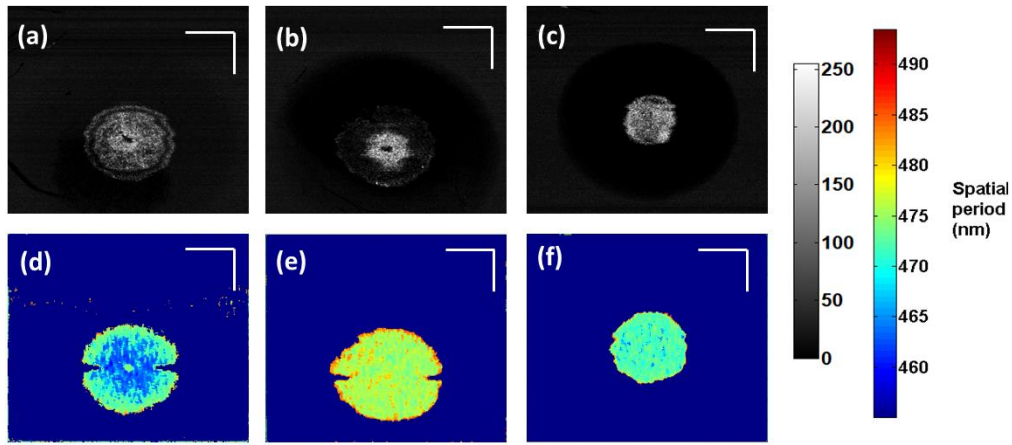

Fig. 4. *Enface* intensity and nsOCT images at a depth of 60 µm. (a), (d) healthy cornea; (b), (e) after alkali induced burn; (c), (f) injured cornea on 7[th] day. Scale bars – 500 µm.

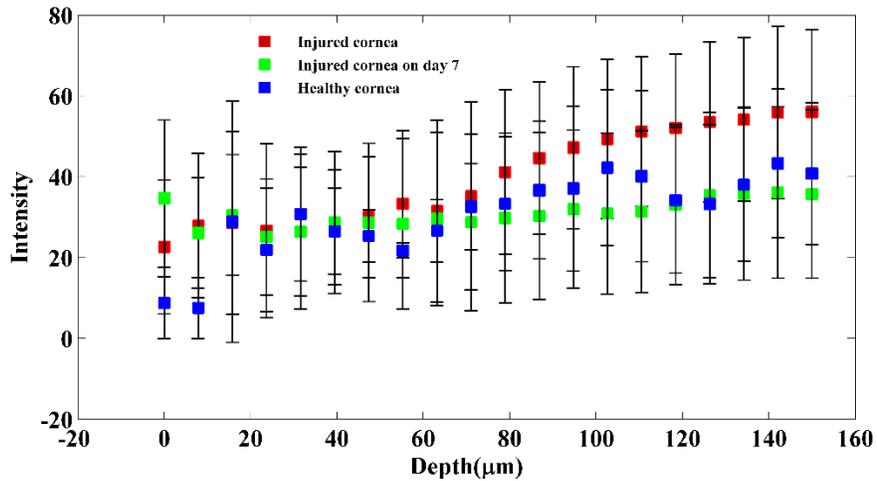

Fig. 5. Plot showing OCT intensity values across the corneal depth between the healthy group, immediate phase group and acute reparative phase group (mean ± std. deviation).

To compare the structural integrity of the cornea between the healthy, immediate phase and acute reparative phase of the injury, dominant spatial period across the corneal depth for all the three groups is plotted in figure 6. Statistical analysis using paired t-test (5 samples per group) shows significant difference ($p < 10^{-10}$) in spatial period changes over the corneal depth between the healthy group, immediate phase group and acute reparative phase groups.

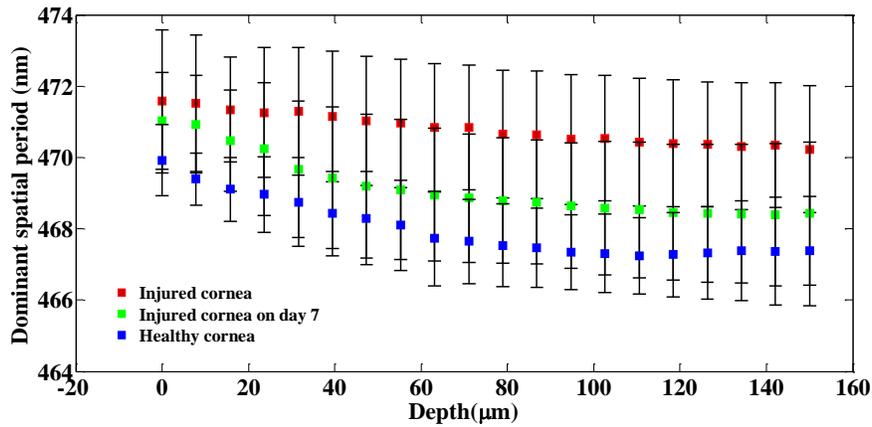

Fig. 6. Plot showing dominant spatial period across the corneal depth between the healthy group, immediate phase group and acute reparative phase group (mean ± std. deviation).

Figure 6 indicates that at the onset of an alkali injury, the spatial period of the structures within the cornea increases at all depths. This is in accordance with the fact that alkali injury penetrates deep into the cornea. The exact reason for this increase in spatial period is not known, however it must be linked to the changes occurring within the collagen matrix of the stroma and due to swelling within the cornea upon the activation of keratocytes within the stroma.

From figure 6, it can be observed that during the acute reparative phase of alkali induced corneal injury, the spatial period of the structures at all depths within the cornea tends to reduce compared to the cornea in the immediate phase group. Also, it is to be noticed that, the most prominent reduction in spatial period occurs within the corneal stroma at depths 50 – 150 µm. These results also indicate that there are significant changes happening within the collagen matrix of the stroma during the acute reparative phase. From figure 6, it can be observed that during the acute reparative phase of the injury, the structural spatial period of the injured cornea across all depths is higher than that of healthy group, however, follows a consistent pattern similar to that of the healthy group. From figures 4 and 6, it is evident that nsOCT is able to capture the nanoscale structural changes within the cornea during the wound healing process *in vivo* thus enabling nsOCT to be a powerful processing method sensitive to nanoscale structural changes within the sample of interest. Also, from figure 6, one can observe the spatially dependent structural periodicity within the cornea at increasing depths in addition to the temporal change. These structural periodicity in different layers of the cornea may be better studied by nsOCT approach using an ultra-high resolution OCT system centred around 800 nm. In the present study, nsOCT algorithm was implemented in Matlab (Mathworks ,version 2014) and takes 30s to process a single nsOCT B- scan using a desktop PC ( DELL Precision T7500, Intel Xenon E5645 2.40 GHz, 12 GB RAM).

In order to calculate the physical spatial periods, we have used an average refractive index value of 1.376 within the cornea. Though the individual corneal layers have different refractive indexes that varies from 1.400 at the epithelium to 1.373 at the endothelium [36,37], the relative changes in refractive index between these layers is less than 0.03, for which the spatial period calculated changes by 10 nm. Previous literature suggests that following alkaline injury, corneal hydration increases [38-40], thereby decreasing the corneal refractive index [39,41]. As alkali injury causes a reduction in average corneal refractive index, the spatial period calculated by nsOCT will be further increased. For simplicity of calculations, the average value of refractive index of a healthy cornea is used in our analysis. Since the purpose of this study was to detect the structural changes within cornea following injury and healing, and not the accurate estimate of the physical spatial period of the structures within each layer of the cornea, mapping the spatial period values by substituting the corresponding refractive index values at each depth within the cornea do not affect the obtained results. This is a proof of concept study to demonstrate the potential of nsOCT to detect structural changes within the cornea following alkali injury and healing, and we have not considered the effect of refractive index changes with wavelength. In the appendix provided along with this manuscript, we have also provided optical spatial period changes (considering refractive index, $n =1$) within the cornea following alkali injury and subsequent healing (figures 9 and 10). The optical spatial period changes are independent of changes in refractive index within biological tissue.

Currently, stained histology imaging is the gold standard to assess micro/nanoscale structural changes within a biological tissue. Thus, besides the nanoscale results obtained using nsOCT, it would be also interesting to analyse the structural changes at microscale using histology sections of the cornea before and after the injury.

Figure 7 (a) and (b) show representative histology sections of a healthy cornea and an injured cornea on day 7 post injury respectively. As seen from figure 7 (a), the healthy cornea is characterized by intact epithelium, and well-arranged collagen fibres within the stroma. However, histological examination of the cornea on day 7 post injury reveals vacuolization of the surface layer of the epithelium along with degenerative changes in the stroma.

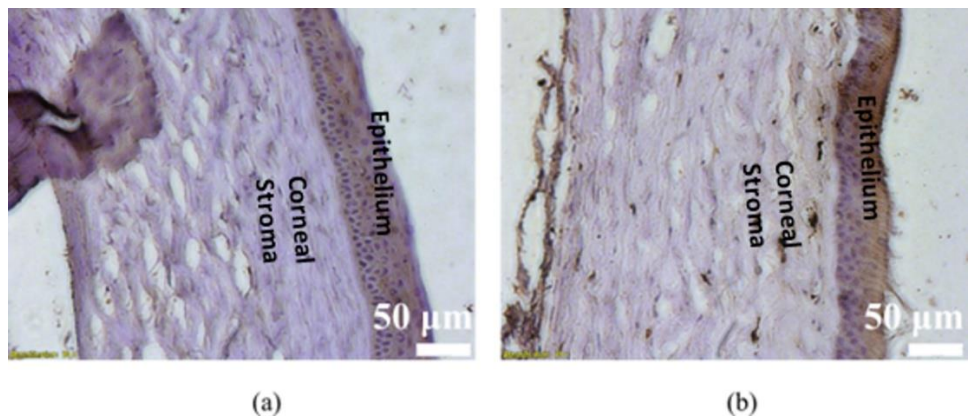

Fig. 7. Corneal histological sectioning (a) healthy cornea (b) injured cornea on day 7; Scale bars – 50 µm.

Furthermore, we hypothesize that to validate nsOCT results with the histological sectioning, spatial frequency changes of the structures within the corneal histology images can be analysed. Recent studies have used multimodal imaging techniques to correlate structural changes within cornea to histological sectioning [42,43]. In order to analyse the spatial frequency profiles from the histology images, we analysed thirty profile lines across the histology cross-section (from the epithelium to the stroma) in a given region and Fourier transforms of these profiles were calculated to obtain the spectrum of spatial period distribution of the structures. From the Fourier spectrum of the profiles, median spatial frequency/period (*msf*) of the spectrum was calculated according to Eq. 4. The *msf* was calculated for thirty profiles in a given region and averaged to give the averaged *msf* to indicate changes within the periodicity of the structures within the cornea pre and post corneal injury.

$$msf = \frac{\sum_{n=1}^{N} f(n)x(n)}{\sum_{n=1}^{N} x(n)} \quad (4)$$

where *x(n)* represents the magnitude of the spectrum at the frequency *f(n)*.

Figures 8(a) and (b) show representative spectra of the line profiles along with the median spatial period values for healthy cornea and injured cornea respectively. It can be observed that the median spatial period tends to move towards the right end of the spectrum for the injured cornea indicating an increase in the spatial period of the structures within cornea post injury. This observation supports our results obtained using the nsOCT technique. For statistical comparison between the median spatial period values of the pre and post injury corneal histology sections, ten sections from each group were analysed. Unpaired t-test results (sample – 10 histology sections per group) show statistical significance between the two groups with $p < 0.05$. Box plot showing the distribution of mean spatial periods of healthy cornea and cornea on day 7 post injury calculated from the histology line profile is shown in figure 8(c). From figure 8(c), it can be observed that post injury, the spatial period of the submicron structures within the cornea has increased compared to that of a healthy cornea. For the healthy cornea, the mean spatial period distribution is $29.3 \pm 6.8$ (standard deviation) µm and for the injured cornea on day 7, it is $34.6 \pm 5$ (standard deviation) µm. So, the results of the Fourier analysis of the histology images supports results obtained using nsOCT approach. The obtained results are interesting to further explore correlations between structural changes occurring at different scales of imaging.

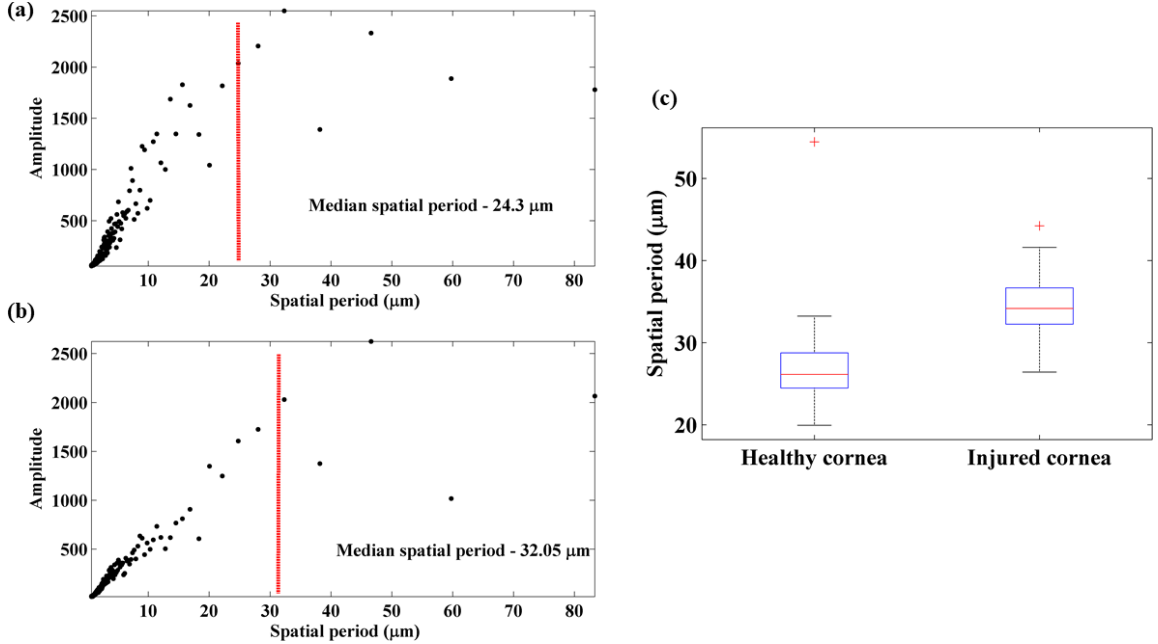

Fig. 8. Spatial period profiles obtained by Fourier transform of histology line profiles (a) healthy cornea (b) injured cornea (c) box plot showing the spatial period distribution of healthy and injured cornea.

## 4. Conclusions

In the present study, along with previous research [15-17, 21], we have elucidated the capability of the nsOCT technique to detect structural changes within the cornea to assess the impact of alkali injury and also to study the wound healing process. nsOCT offers much higher sensitivity to structural changes within the cornea compared to conventional OCT processing. The study reveals that nsOCT is able to detect structural changes with nanoscale sensitivity between healthy cornea, injured cornea and also during the reparative phase of the injury at all depths within the cornea with high statistical significance ($p < 10^{-10}$). Further studies are required to accurately determine the physical spatial period changes taking place within cornea following the injury by considering the refractive index values at each of the layers within cornea.

The method presented offers potential for *in vivo* imaging applications especially in clinical imaging where sensitivity to changes in structure is of significance either to detect the onset of a disease or to evaluate the efficacy of treatment which cannot be obtained from conventional OCT processing. In contrast to phase sensitive OCT, nsOCT images are formed from just a single frame. Further studies are required to assess the suitability of the method described to measure the corneal transparency based on the sensitivity of nsOCT to the structural integrity of collagen network within the stroma. The technique described bridges the gap between high resolution imaging and increased depth of imaging in OCT by enhancing the sensitivity of OCT to nanoscale structural changes within the sample. Further applications of the technique can be used to study morphological changes in biomedical samples, for example, to image progression of cancerous cells and tumours as they are known to undergo nanoscale structural changes within their vicinity long before the manifestation of the disease.

## 5. Appendix

Figure 9 shows the comparison of optical spatial period changes across the corneal depth for the healthy group and the immediate phase group. Statistical analysis using paired t-test (samples 12 per group) shows significant difference ($p < 10^{-7}$) in optical spatial period changes over the corneal depth between the healthy uninjured cornea and the immediate phase group.

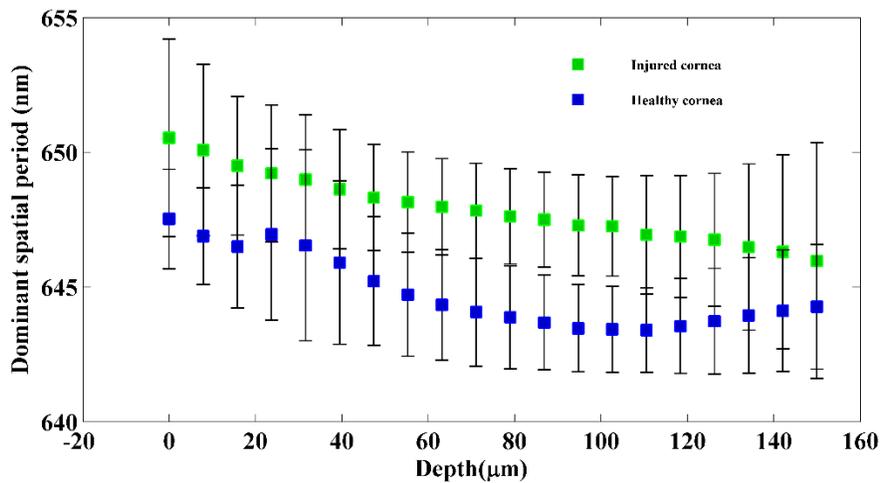

Fig. 9. Averaged dominant spatial period across the corneal depth for the healthy group and the immediate phase group.

Figure 10 shows the comparison of optical spatial period changes across the corneal depth for all the three groups.

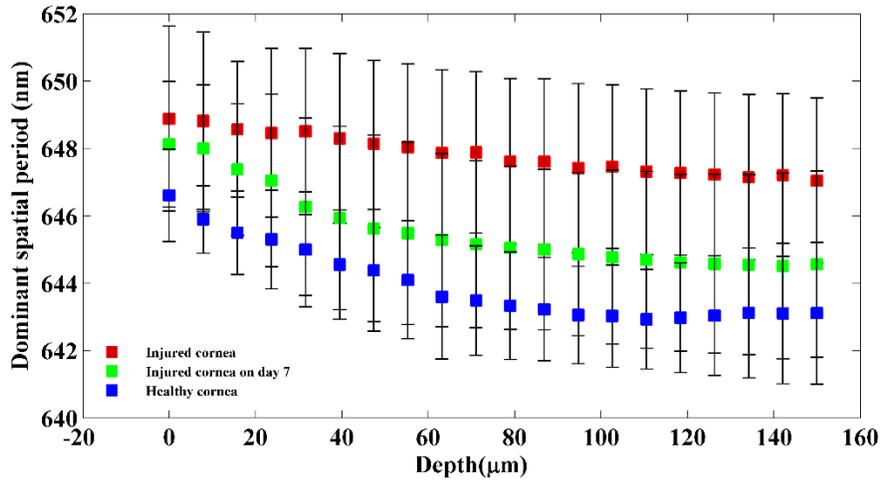

Fig. 10. Plot showing dominant spatial period across the corneal depth between the healthy group, immediate phase group and acute reparative phase group.

Figure 11 shows the plot of physical spatial period changes across the corneal depth for all the three groups together with the individual data points ( all the animals ).

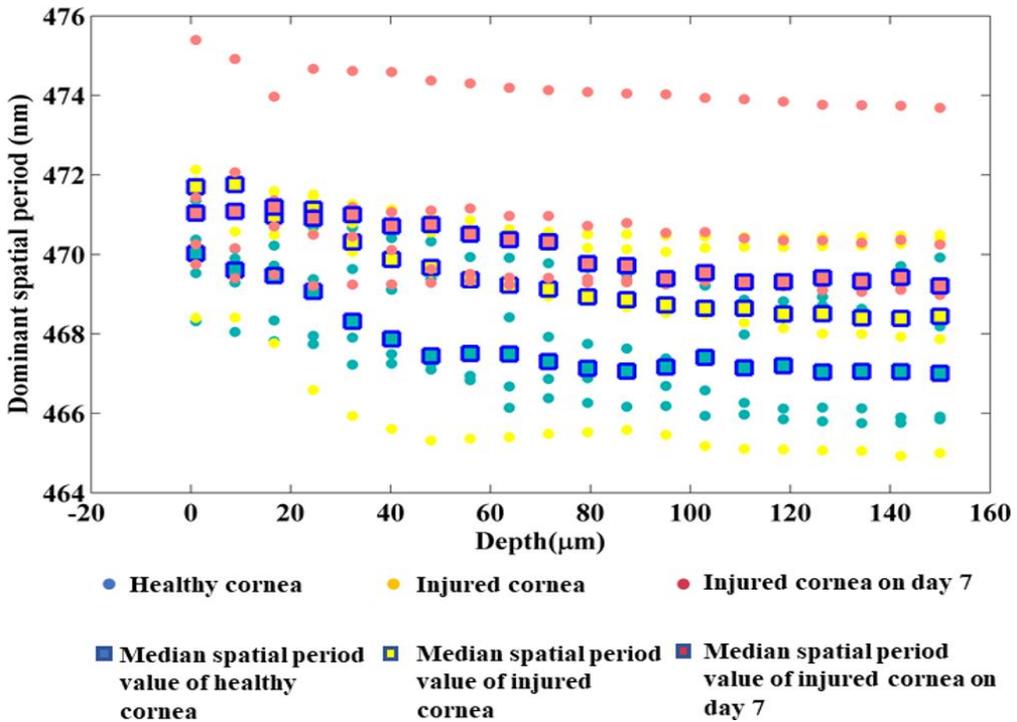

Fig. 11. Plot showing dominant spatial period across the corneal depth between the healthy group, immediate phase group and acute reparative phase group together with the individual data points ( all the animals ).

Table 1 shows the calculations used to find the depth at which most prominent reduction in spatial period occurs during the acute reparative phase. The data used for this table comes from the plot in figure 6 of the manuscript.

Table 1. Tabulated values of percentage reduction of physical spatial changes across the corneal depth between immediate phase and acute reparative phase.

| Depth (µm) | Percentage reduction = ((Immediate phase - Reparative phase)/Immediate phase) x 100 |
|---|---|
| 0 | 0.12 |
| 8 | 0.12 |
| 16 | 0.18 |
| 24 | 0.22 |
| 32 | 0.34 |
| 40 | 0.36 |
| 48 | 0.39 |
| 56 | 0.40 |
| 64 | 0.42 |
| 72 | 0.40 |
| 80 | 0.40 |
| 88 | 0.41 |
| 96 | 0.40 |
| 104 | 0.40 |
| 112 | 0.40 |
| 1201 | 0.41 |
| 128 | 0.41 |
| 136 | 0.41 |
| 144 | 0.41 |
| 152 | 0.38 |

# 6. Funding, acknowledgments, and disclosures

## 6.1 Funding

This project has received funding from the European Union's Horizon 2020 research and innovation program under grant agreements No 761214 and 779960. The materials presented and views expressed here are the responsibility of the authors(s) only. The EU Commission takes no responsibility for any use made of the information set out.

## 6.2 Acknowledgements

The authors acknowledge the help provided by Dr Hrebesh Molly Subhash during the experiments.

## 6.3 Disclosures

The authors declare no conflicts of interest.

## 6.4 Ethics statement

This study was approved by Animals Care Research Ethics Committee of the National University of Ireland, Galway. All the experimental procedures were performed in accordance with and authorization from the Health Products Regulatory Authority of Ireland.